\documentclass[sigconf,screen,nonacm]{acmart}

\setcopyright{none}
\acmDOI{}
\acmISBN{}
\acmYear{}
\copyrightyear{}

\date{Preprint — October 2025}

\usepackage{pifont}
\usepackage{colortbl}
\newcommand{\tool}{InvisibleMentor}
\newcommand{\term}{vision-grounded task reflection}
\arrayrulecolor{gray}

\usepackage{listings}
\usepackage{textcomp}

\begin{document}
\title{The Invisible Mentor: Inferring User Actions from Screen Recordings to Recommend Better Workflows}

\settopmatter{authorsperrow=4}
\author{Litao Yan}
\email{ltyan@seas.upenn.edu}
\orcid{0009-0009-5077-354X}
\affiliation{%
  \institution{University of Pennsylvania}
  \city{Philadelphia}
  \state{PA}
  \country{USA}
}

\author{Andrew Head}
\email{head@seas.upenn.edu}
\orcid{0000-0002-1523-3347}
\affiliation{%
  \institution{University of Pennsylvania}
  \city{Philadelphia}
  \state{PA}
  \country{USA}
}

\author{Ken Milne}
\email{kenmilne@microsoft.com}
\affiliation{%
  \institution{Microsoft}
  \city{Redmond}
  \state{WA}
  \country{USA}
}

\author{Vu Le}
\email{levu@microsoft.com}
\orcid{0000-0003-3727-3291}
\affiliation{%
  \institution{Microsoft}
  \city{Redmond}
  \state{WA}
  \country{USA}
}

\author{Sumit Gulwani}
\email{sumitg@microsoft.com}
\orcid{0000-0002-9226-9634}
\affiliation{%
  \institution{Microsoft}
  \city{Redmond}
  \state{WA}
  \country{USA}
}

\author{Chris Parnin}
\email{chrisparnin@microsoft.com}
\orcid{0000-0001-6182-815X}
\affiliation{%
  \institution{Microsoft}
  \city{Redmond}
  \state{WA}
  \country{USA}
}

\author{Emerson Murphy-Hill}
\email{emerson.rex@microsoft.com}
\orcid{0000-0003-3921-9416}
\affiliation{%
  \institution{Microsoft}
  \city{Redmond}
  \state{WA}
  \country{USA}
}

\begin{abstract}
Many users struggle to notice when a more efficient workflow exists in feature-rich tools like Excel. Existing AI assistants offer help only after users describe their goals or problems, which can be effortful and imprecise. We present \tool{}, a system that turns screen recordings of task completion into \term{}. It detects issues such as repetitive edits and recommends more efficient alternatives grounded in observed behavior. Unlike prior systems that rely on logs, APIs, or user prompts, \tool{} operates directly on screen recordings. It uses a two-stage pipeline: a vision-language model reconstructs actions and context, and a language model generates structured, high-fidelity suggestions based on the reconstructed workflow. In evaluation, \tool{} accurately identified inefficient workflows, with participants finding its suggestions more actionable, tailored, and helpful for learning and future improvement than those of a prompt-based spreadsheet assistant.
\end{abstract}



\begin{teaserfigure}
  \includegraphics[width=\textwidth]{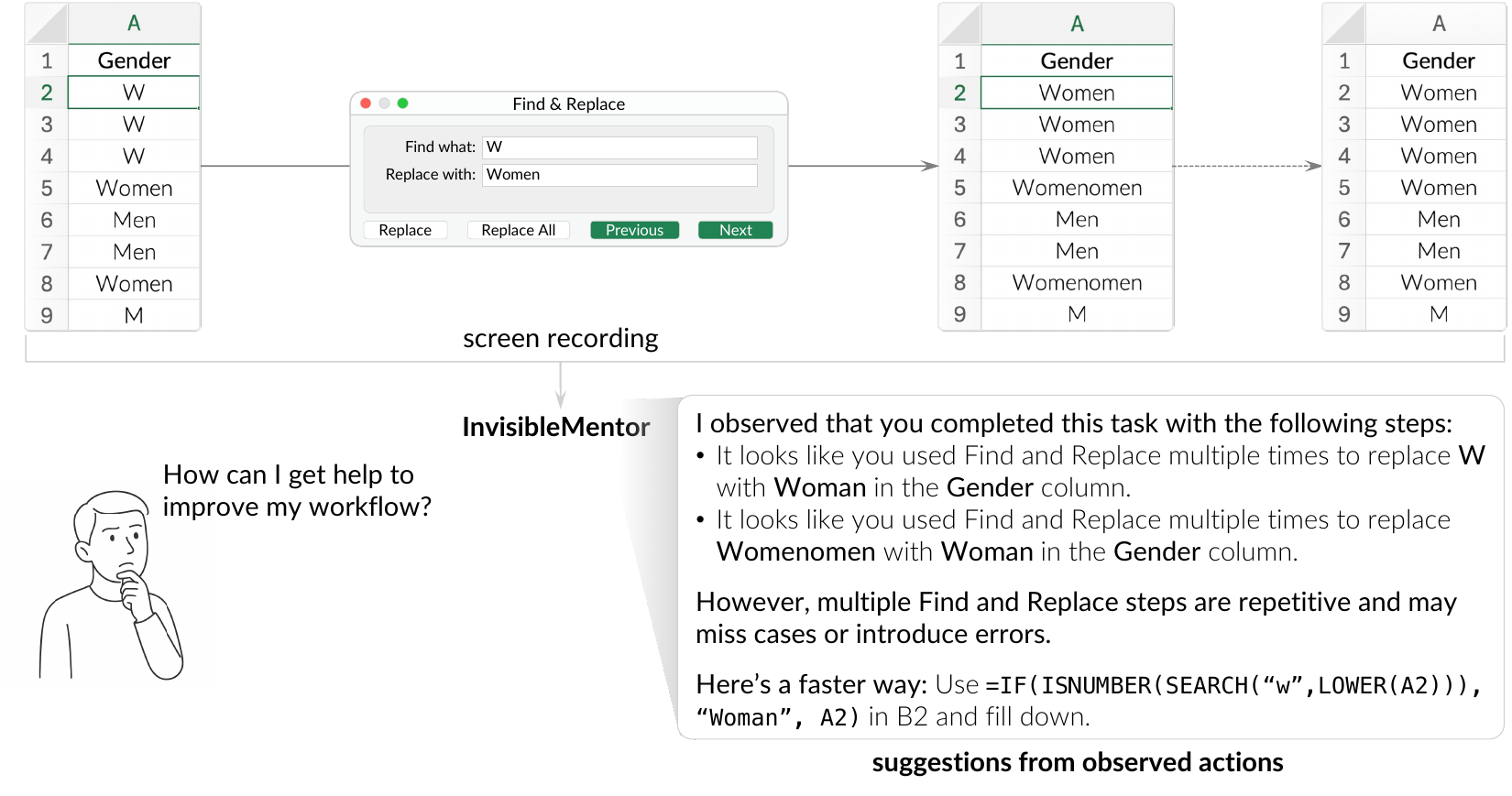}
  \vspace{-4ex}
  \caption{\tool{} transforms ordinary screen recordings into more than just sequences of user actions. It interprets the user's workflow, identifies inefficiencies such as repetitive Find \& Replace steps that introduce errors, and generates concrete alternatives grounded in what the user just did. By learning from observed behavior, rather than from logs or explicit prompts, \tool{} offers structured, high-fidelity suggestions on how workflows can be improved.}
  \label{fig:teaser}
  \vspace{2ex}
\end{teaserfigure}

\maketitle
\section{Introduction}
\label{sec:introduction}
Complex software applications such as Microsoft Excel provide users with an extensive array of features and enable sophisticated workflows that can seamlessly integrate these capabilities to support efficient and effective work.
Yet, most users tend to utilize only a limited subset of these tools and workflows. 
For example, Figure 1 (top) illustrates a participant in our study employing the \textit{Find and Replace} feature to normalize a data column containing gender information---a method that is both tedious and prone to error. 
Advanced Excel features such as formulas or Flash Fill could greatly facilitate such tasks, if only users were aware of how and when to employ them.

One of the most effective ways users learn complex software is through over-the-shoulder learning~\cite{twidale2005otsl} or peer observation~\cite{murphy2015users}, where a more experienced user notices inefficiencies and offers timely, grounded advice.
But such learning moments are rare, limited by chance, physical proximity, and the peer's expertise~\cite{qing2025}.
How might we increase the frequency and accessibility of these valuable learning experiences?

To do so, we propose a technique that we call \textit{\term{}},
where generative artificial intelligence assumes the role of the observing peer. Our system, \tool{}, embodies this concept, following a two-phase pipeline for recommending improved workflows in Excel tasks (Figure~\ref{fig:pipeline}). In the first phase, a vision-language model (VLM) analyzes screen recordings to reconstruct a structured representation of the user's task, including action sequences (e.g., inserting rows, editing formulas) and spreadsheet changes. Unlike prior systems, \tool{} requires no system logs, internal APIs, or explicit input. In the second phase, a textual language model generates structured, high-fidelity recommendations, offering step-by-step guidance tailored to the user's actual workbook. This design minimizes the effort required to recall and articulate workflow details, while enabling personalized and context-specific suggestions. Our goal in building this pipeline is to explore the feasibility and value of \term{}, and to inform the design of future AI assistants.

This approach is attractive for three reasons. First, by recovering both action sequences and spreadsheet context, the system can generate precise, context-specific recommendations. Second, by observing entire workflows, it can uncover missed opportunities, such as overlooked features or reordering steps. Third, by shifting the burden of articulation from user to system, \tool{} makes help accessible without requiring prompt crafting or prior intent.

We evaluate \tool{} through a benchmark and a lab study. In the benchmark, the vision-language model achieved over 90\% accuracy on recovering 14 common spreadsheet actions across 25 real sessions. In a within-subjects study with 20 participants, users significantly preferred \tool{}'s suggestions over those from a prompt-based spreadsheet assistant. They found the suggestions more useful, easier to understand, and less effortful to apply. \tool{} also helped participants notice inefficiencies, identify mistakes, and feel confident trying alternative workflows.

This paper makes the following contributions:
\begin{itemize}
\item Vision-grounded task reflection, a new interaction mode that offers improvement suggestions based on observed user behavior, without requiring prompts, instrumentation, or intent specification.
\item \tool{}, a two-phase system that recovers workflows from screen recordings and generates structured, high-fidelity suggestions to improve future workflows.
\item A benchmark and user study showing that \tool{} accurately detects inefficient workflows and supports better user reflection, learning, and improvement.
\end{itemize}
\section{Background and Related Work}
\subsection{Feature Discoverability and Action-Based Suggestions}
\label{sec:featureSuggestions}

Users of feature-rich applications routinely use only a small subset of available functionality due to deep menus, terminology mismatches, and lack of guidance~\cite{Nielsen1993,Mackay1990,Findlater2004}. In spreadsheets, even experienced professionals tend to rely on basic arithmetic operations and manual cell references, underutilizing advanced features such as macros or pivot tables~\cite{bradbard2014spreadsheet,srinivasa2021spreadsheet}. Murphy-Hill et al.\ show that users discover new tools infrequently and largely through non-systematic channels, underscoring a persistent discoverability gap~\cite{murphy2015users}. Similarly, Kohlhase et al.\ find that spreadsheet readers often ``miss the context'' intended by authors, making it hard to recognize relevant features~\cite{kohlhase2015context}. 

Spreadsheets especially highlight this gap: users often rely on repetitive manual steps or overly complex formulas when simpler built-in functions would suffice~\cite{Chalhoub2022,srinivasa2021spreadsheet}. Even when tools exist, false affordances and ambiguous icons can mislead users as they scan for something that ``looks right''~\cite{lafreniere2017falseaffordance}. Recent work further shows that discoverability barriers are not only technical but also social: Xia et al.\ identify how social norms and conflicted evaluations of spreadsheet expertise can limit knowledge sharing, underscoring the need for sociotechnical approaches to feature adoption~\cite{xia2025barriers}.

To improve discoverability, prior work has proposed a range of approaches. Command search reduces hunting costs but fails when user terminology diverges from product vocabulary~\cite{Fourney2011}. In-situ help provides guidance at the point of use. For example, LemonAid presents crowd-sourced Q\&A linked to UI elements, allowing users to access relevant solutions without leaving the application~\cite{Chilana2013LemonAid}, and ToolClips embeds micro-videos into tooltips for just-in-time instruction~\cite{Grossman2010ToolClips}. While these methods lower navigation costs, they often require users to initiate search and are not grounded in their current context or goals.

In contrast, systems that use user actions generate context-aware suggestions without explicit prompting. Early adaptive interfaces, such as Eager~\cite{Cypher1993Eager}, inferred repetitive behaviors and proactively offered automation. Similarly, adaptive menus in Excel dynamically reordered commands based on usage patterns~\cite{Findlater2004}, showing that incorporating behavioral signals can improve efficiency while reducing cognitive load. Usage-based recommendation systems, such as CommunityCommands~\cite{Matejka2009CommunityCommands}, Patina~\cite{Matejka2013Patina}, and IDE fluency assistants~\cite{murphyhill2012fluency}, analyze historical usage patterns to uncover underused commands and highlight expert practices. Complementing these, tools like DiscoverySpace~\cite{Fraser2016} and Ambient Help~\cite{Matejka2011} instead provide real-time, context-aware suggestions based on the user's current task state or recent actions. In contrast, AdaptableGIMP~\cite{AdaptableGIMP} allows users to manually browse and apply crowd-contributed workflows rather than proactively recommending actions.

More recent work advances these ideas using data-driven prediction. Wang et al.~\cite{Wang2018} analyzed millions of interaction logs and tutorial videos to recommend workflow steps matching current tasks. Nambhi et al.~\cite{Nambhi2019} inferred user intent from analytics tool event sequences, enabling timely suggestions when users struggle. In software engineering, Bulmer et al.~\cite{Bulmer2018} applied machine learning to large-scale logs of developer interactions in the IDE to predict likely next actions, supporting intelligent code-editing assistance. However, most of these approaches rely on structured logs or application instrumentation, limiting where they can be deployed. Our system instead infers task context and structure solely from visual evidence, enabling broader applicability without requiring internal access to software.

\subsection{Recovering Task Representations from Screens}
\label{sec:recover-task}

Recent work advances screen and UI understanding by recognizing elements and recovering layout structure. Vision-based models like ScreenAI~\cite{screenai2024}, MobileVLM~\cite{mobilevlm2024}, and ILuvUI~\cite{iluvui2025} support element targeting, instruction following, and inter-screen understanding. Other methods reconstruct user actions and workflows from screencasts~\cite{seehow2023,codeextract2023}, recovering fine-grained interaction sequences, and action-aware models map visual changes to operations~\cite{actionnet2019}. Alahmadi et al.\cite{uiscreens2020} extract representative UI screens from tutorial videos to segment long sessions into meaningful, task-relevant states. Similarly, analyses of ``watch me code or work'' videos have shown that detailed logs of user activity can be mined to understand task progress and hurdles\cite{watchmecode2018}.

Some systems capture detailed records of application usage through instrumentation, such as undo logs~\cite{akers2009undo} or sequences of IDE commands~\cite{murphyhill2012fluency}. Others infer intent in a less intrusive way by observing aggregate behavior patterns. For example, repeated scrolling or clicking activity can reveal which parts of an interface attract attention~\cite{deline2005wear}, and resumption cues can indicate how users re-engage with tasks after an interruption~\cite{parnin2010resumption}.These approaches help recover implicit structure, task progress, and user struggles, offering new lenses into user workflows without requiring manual labeling.

Domain-specific analyses show how task representations can be reconstructed in particular contexts. In spreadsheets, reasoning over cell contents, ranges, headers, and selection history is key to understanding user intent~\cite{kohlhase2015context,srinivasa2021spreadsheet}. Other work demonstrates how UI interactions can be extracted from videos and even translated across applications, as in ShowMeHow~\cite{ramesh2011showmehow}, highlighting the promise of video-driven instruction recovery. These representations offer a promising foundation for AI assistants aiming to generate recommendations grounded in real user activity. 

Recent work has begun applying vision-language models (VLMs) to infer user behavior directly from screen recordings. Sharingan~\cite{chen2024sharingan} uses a VLM to label GUI actions in desktop video recordings, demonstrating the feasibility of extracting structured action sequences without application instrumentation. Similarly, ScreenLLM~\cite{xu2025screenllm} transforms screen recordings into a compact stateful schema describing UI changes and uses a textual LLM to interpret or predict user actions. In mobile domains, V2S+~\cite{bernal2023v2splus} extracts replayable gesture scripts from Android screen videos using computer vision techniques. While promising, these systems primarily focus on action reconstruction or replay, rather than providing user-facing feedback or personalized guidance.

\subsection{AI Assistants for Task-Focused Recommendations}
\label{sec:ai-recommendations}

Chat-based assistants like Excel Copilot allow users to issue commands via natural language. However, studies show that non-experts struggle to articulate goals, recall terminology, and provide adequate prompts~\cite{Zamfirescu2023,Khurana2025}. Even when functional, such assistants often lack grounding in the user's active artifact or recent steps, leading to vague or irrelevant responses~\cite{Khurana2024}. These limitations highlight the need for alternatives that observe what users do, rather than rely solely on what they say.

As an alternative, research explores recommendation methods that infer intent from user actions, anchoring suggestions in ongoing work rather than explicit prompts. Example-driven programming systems infer operations from user edits~\cite{Singha2024,Gulwani2012CHI,Singh2016POPL}, while crowd-powered systems recommend fixes for compiler errors based on prior user behavior~\cite{hartmann2010helpmeout}. Mixed-initiative tools~\cite{Chen2024} propose action sequences but let users approve or modify them, balancing automation with control. Recent systems extend these directions: TableTalk uses a language agent to scaffold spreadsheet construction incrementally, yielding higher-quality outcomes with lower effort~\cite{liang2025tabletalk}, and large-scale macro mining demonstrates how analyzing records of user actions can provide reusable workflows and automation opportunities~\cite{zhang2024macroMining}. Together, these approaches suggest that assistants based on users' ongoing activity can deliver more relevant and timely guidance, avoiding the need for elaborate prompting and tying recommendations directly to what users are actually doing.
\section{System Design}
\label{sec:system}

We designed \tool{} to minimize user effort and maximize contextual relevance. Below, we articulate four goals that shaped the system, followed by an overview of its architecture and implementation.

\subsection{Goals}
\paragraph{G1. Translate visual representations into semantically rich representations without specialized instrumentation.}
Many support systems rely on internal logs or APIs to detect context and offer help (Section~\ref{sec:recover-task}). While effective, these methods constrain where guidance can be offered, requiring access that is often unavailable in commercial or cross-application settings. In contrast, our goal is to extract rich, contextual representations directly from screen recordings, enabling guidance across tools with limited or no internal access.

\paragraph{G2. Provide structured, high-fidelity suggestions grounded in user behavior.}
The suggestions should not only be helpful but also intelligible and trustworthy. To achieve this, these suggestions should explicitly show what users actually did, not abstract templates, and should explain why a workflow might be inefficient, explaining missed opportunities. They should also be concrete and specific, with sufficient detail to be immediately actionable.

Together, G1 and G2 distinguish our work from prior approaches: we offer richly contextualized guidance without internal instrumentation by interpreting screen activity alone. These goals are complemented by additional design considerations commonly found effective:

\paragraph{G3. Discover unspoken help opportunities by modeling user behavior directly from observation.}
Most systems wait for users to ask for help (Section~\ref{sec:ai-recommendations}), but users often fail to recognize inefficiencies. As a result, help opportunities go unnoticed and underutilized. Instead, the system should proactively infer opportunities for improvement from observed behavior, like a knowledgeable peer who notices when a task could be done more easily.

\paragraph{G4. Deliver guidance shortly after task completion to promote reflective learning}
Mid-task suggestions risk interrupting focus and increasing cognitive load. By offering help just after a task ends, when memory is fresh and attention is available, the system can better support reflection, reinforce strategies, and encourage long-term learning.

\begin{figure*}
    \centering
    \includegraphics[width=\linewidth]{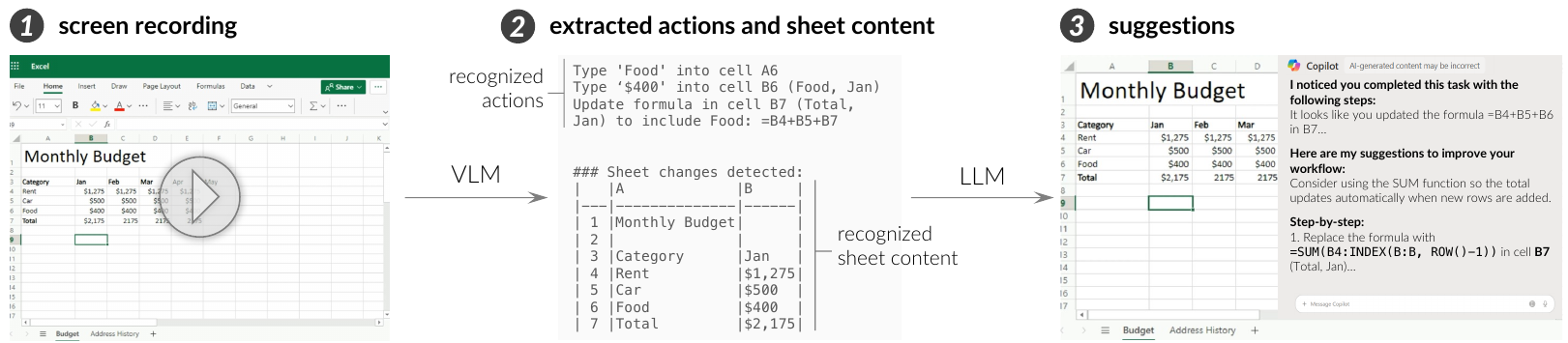}
    \vspace{-2em}
    \caption{\tool{}'s pipeline for generating suggestions from a screen recording. \textmd{The system operates in two phases. (\textcolor[HTML]{595959}{\ding{202}}) A vision-language model (VLM) processes screen recordings sampled every 5 seconds to extract structured task representations, including user actions and spreadsheet context (\textcolor[HTML]{595959}{\ding{203}}). These representations are grouped into workflows and passed to a language model (LLM), which analyzes them to identify inefficiencies and generate actionable suggestions. Each suggestion includes a sequence of inefficient workflow, a rationale, a step-by-step suggestion (\textcolor[HTML]{595959}{\ding{204}}).}}
    \label{fig:pipeline}
    \Description[]{}
\end{figure*}

\subsection{Implementation}
We implemented \tool{} as a two-phase pipeline that combines visual interpretation of user activity with textual reasoning for recommendations (Figure~\ref{fig:pipeline}). The first phase uses a vision-language model (VLM) to capture a structured representation of the user's task from screen recordings (Figure~\ref{fig:pipeline}, \textcolor[HTML]{595959}{\ding{202}}), while the second phase employs a large language model (LLM) to generate grounded and actionable suggestions. We describe each phase in detail below.

\subsubsection{Phase 1: Capturing Task Context with a VLM}

The first phase constructs a representation of the user's task based on visual observation (G1). As the user works in Excel, the system samples one screen frame every five seconds to capture a lightweight sequence of their on-screen actions. This cadence balances temporal coverage with efficiency: it is frequent enough to capture meaningful actions, such as formula edits, filter changes, or table creation, while avoiding redundant frames that increase cost and latency.

\paragraph{Task representation.}
Each frame encodes the spreadsheet state and interface context, including selected cells, visible formulas, column headers, named tables, and open panes. To analyze these frames, we use GPT-4.1\footnote{All experiments used the Azure OpenAI Service deployment of GPT-4.1 (April 2025 release).}, which supports both text and image input and provided state-of-the-art image understanding during the period of development. The model produces a structured representation of user activity with two elements:  
(a) \textit{Action sequence}, a list of low-level user actions (e.g., editing a formula, applying a filter, switching sheets); and  
(b) \textit{Contextual state}, the spreadsheet content and layout at each moment, such as selected ranges, cell values, and interface elements (Figure~\ref{fig:pipeline}, \textcolor[HTML]{595959}{\ding{203}}).  

Because the model's context window can accommodate only about 20 frames, roughly 100 seconds of video at our chosen interval, a long recording cannot be processed in a single pass.\footnote{While GPT-4.1 supports up to 1M tokens in Azure, practical limits depend on deployment type and prompt overhead. In our configuration (128k managed deployment, structured prompts, and metadata per frame), the effective budget allowed for about 20 1080p frames per call.} To address this limitation, we divide each recording into up to three temporal segments, each capturing a portion of the task. These segments are processed in parallel to reduce post-task latency. Within each segment, we further divide the frames into batches of 20 frames, each covering approximately 100 seconds of video. These batches are fed sequentially into separate VLM calls. This hierarchical strategy enables the system to recover a complete task trace from longer recordings, while respecting the model's input limits and preserving continuity within and across segments.

\paragraph{Pilot testing and parameter choices.}
To select appropriate parameters, we conducted a pilot test using the \textit{Europe Bike Store Sales} dataset, in which the authors performed three representative spreadsheet tasks. The entire session was recorded, producing a one-hour video for VLM testing. We compared different sampling intervals and segmentation strategies. At a 1-second interval, the VLM repeatedly detected near-duplicate actions because a single user action typically spanned several seconds; this redundancy inflated processing time and yielded little additional information. In contrast, a 10-second interval often skipped important, transient behaviors, such as brief pop-up messages or rapid double-clicks to apply formulas. Balancing coverage with efficiency, we chose a 5-second interval, which avoided both excessive duplication and missed actions.

We also compared processing times with and without segmentation. Without segmentation, processing a one-hour video at a 5-second interval required 6 minutes, and nearly 41 minutes at a 1-second interval---approaching the duration of the video itself. With segmentation into three parallel chunks, processing time dropped to around 2 minutes at the 5-second interval. We discuss the limitations of this sampling strategy in Section~\ref{sec:limitations}.

\subsubsection{Phase 2: Generating and Delivering Recommendations with an LLM}

The second phase generates structured and high-fidelity suggestions from the extracted task representation (G2 and G3). Whereas Phase 1 relied on GPT-4.1's vision capabilities, Phase 2 uses the same model in textual mode, taking structured actions and contextual states as input under a structured zero-shot prompt (see Appendix~\ref{sec:appendix-prompts}).

\paragraph{Recommendation generation.}
The prompt instructs the model to group related actions into workflows, mark those that are suboptimal, and return at most three prioritized recommendations. Each recommendation contains the relevant actions and a rationale for why they are inefficient, together with a concrete alternative (including explicit steps or formulas) and a brief ``benefit'' statement comparing before and after (Figure~\ref{fig:pipeline}, \textcolor[HTML]{595959}{\ding{204}}).

Rather than streaming partial results, \tool{} processes the set of observed steps from Phase 1 before generating recommendations. This design enables the LLM to detect not only local inefficiencies but also higher-level repetitive patterns. For example, while a sequence of copy–paste operations may be inefficient at the action level, broader repetition often occurs at the workflow level, such as repeatedly cycling through data cleaning, processing, and analysis. Within each cycle, individual actions may appear reasonable, but when viewed together they reveal opportunities for more efficient workflows. By reasoning over the full sequence of observed user actions, the system can thus provide suggestions that address both fine-grained operations and broader workflow structures.

\begin{figure*}
    \centering
    \includegraphics[width=\linewidth]{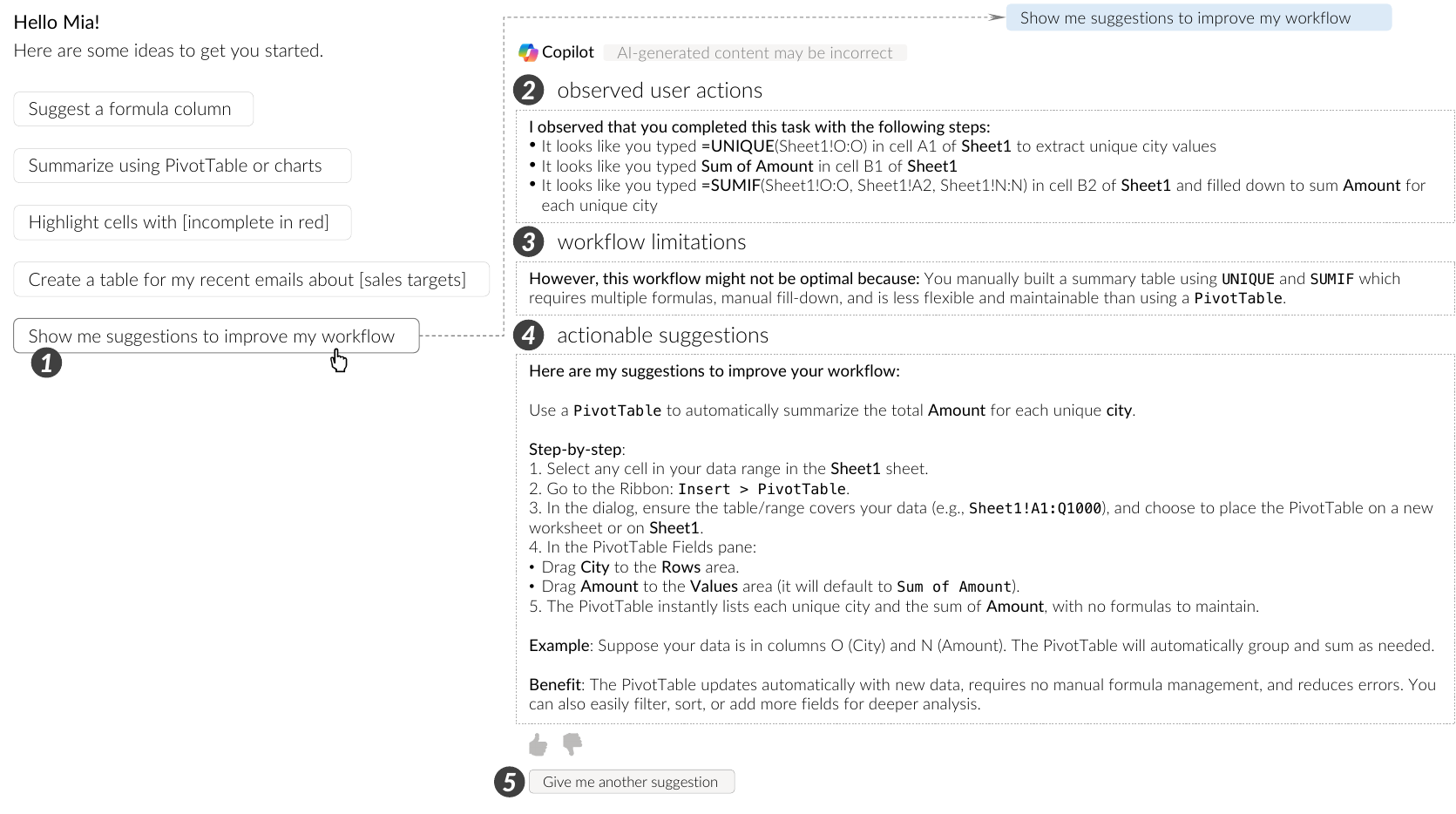}
    \vspace{-2em}
    \caption{User interface of a spreadsheet assistant that provides structured workflow guidance. \textmd{The assistant appears in a task pane alongside the spreadsheet, following the familiar layout of Excel Copilot to minimize design variance that could influence user study outcomes. The interface consists of five key components. (\textcolor[HTML]{595959}{\ding{202}}) Prompt ideas offer users high-level suggestions to initiate help-seeking. Selecting an idea appends a message to the conversation and triggers a structured assistant response. To streamline access to suggestions, we replaced the original fifth idea with a dedicated entry point for requesting recommendations. The remaining components are rendered dynamically based on model output: (\textcolor[HTML]{595959}{\ding{203}}) observed user actions, summarizing recent spreadsheet activity; (\textcolor[HTML]{595959}{\ding{204}}) workflow limitations, explaining potential inefficiencies; (\textcolor[HTML]{595959}{\ding{205}}) actionable suggestions, presenting step-by-step improvements; and (\textcolor[HTML]{595959}{\ding{206}}) a ``give me another suggestion'' button, allowing users to request alternatives.}}
    \label{fig:interface}
    \Description[]{Stacked bar chart showing participant preference, effort, and privacy confidence comparing TrailBlazer with a baseline system.}
\end{figure*}

\subsubsection{Interface}
The interface (Figure~\ref{fig:interface}) appears after task completion (G4). It is displayed in a task pane alongside the spreadsheet, following the familiar layout of Excel Copilot. This design minimizes interface-related confounds, helping ensure that observed effects in the study reflect the system's recommendations rather than novel UI elements.\footnote{See our user scenario in Appendix~\ref{sec:scenario}.}

\paragraph{Prompt ideas as entry points}
To reduce visual differentiation from the original Excel Copilot interface, we replaced the fifth idea card with a dedicated entry point labeled ``Show me suggestions to improve my workflow'' (Figure~\ref{fig:interface}, \textcolor[HTML]{595959}{\ding{202}}).

\paragraph{Observed user actions}
The assistant summarizes the user's recent activity in natural language (Figure~\ref{fig:interface}, \textcolor[HTML]{595959}{\ding{203}}), highlighting workflows identified as potentially inefficient (G2). Each description reflects a real sequence of user actions, e.g., building a summary table using UNIQUE and SUMIF, grouped into meaningful task units. These summaries serve two purposes. First, they help users recall and recognize inefficiencies that may have gone unnoticed during task completion, supporting accurate memory alignment. Second, by explicitly surfacing these patterns, the assistant builds credibility for subsequent suggestions, allowing users to judge whether its interpretation aligns with their memory and experience.

\paragraph{Workflow limitations}
The system explains potential inefficiencies in the observed workflow (Figure~\ref{fig:interface}, \textcolor[HTML]{595959}{\ding{204}}), such as using multi-step formulas where built-in features would suffice. These explanations help users understand not just what could change, but why.

\paragraph{Actionable suggestions}
Recommendations are presented as concrete, step-by-step instructions (Figure~\ref{fig:interface}, \textcolor[HTML]{595959}{\ding{205}}), tailored to the user's observed workflows (G2). Each suggestion explicitly names the relevant features (e.g., Get Data, Merge Queries) and provides exact Ribbon paths or shortcuts for accessing them. When applicable, the system also includes output examples to help users verify their results. Suggestions are designed to be executable within the spreadsheet environment, even for advanced features like Power Query or PivotTables. Each ends with a concise benefit statement that compares the original workflow to the recommended one, highlighting time savings, reduced errors, or improved maintainability (e.g., ``Original: 4+ steps, Suggested: 2 steps''). This framing encourages adoption by making the value of the recommendation clear and tangible.

\paragraph{Alternatives on demand}
To reduce information overload, the assistant initially presents only the most critical recommendation. Users can request additional suggestions using the ``Give me another suggestion'' button (Figure~\ref{fig:interface}, \textcolor[HTML]{595959}{\ding{206}}), which sequentially reveals other suboptimal workflows identified during analysis. This interaction model enables a lightweight, controlled exploration of the full recommendation set, allowing users to focus on one improvement at a time without becoming overwhelmed.
\section{Technical Evaluation}

To assess the technical feasibility of our approach, we conducted a technical evaluation focused on how accurately a vision-language model (VLM) can recover spreadsheet actions from screen recordings. This evaluation was guided by the following research question:

\smallskip
\emph{RQ1.} How accurately can a VLM identify users' spreadsheet actions from screen recordings?

\subsection{Methodology}
\label{sec:eval1-setup}
\paragraph{Dataset}
To evaluate RQ1, we used an internally collected benchmark dataset of spreadsheet interaction videos. The dataset contains 25 screen-recorded sessions where participants, recruited via PlaybookUX, completed 14 short tasks involving basic yet practical Excel operations, such as inserting rows, editing formulas, switching sheets, and renaming tabs. Each session used an Excel workbook with two pre-filled sheets, each containing 6 to 12 rows of synthetic data.\footnote{See Appendix~\ref{appendix:eval1-details} for a detailed list of tasks and spreadsheet content used in Evaluation~1.} This dataset provides a suitable basis for evaluating the VLM's ability to recognize fine-grained user actions in spreadsheet environments. We were not involved in its original collection but used it for evaluation.

\paragraph{Procedure}
Before evaluation, we preprocessed all videos by cropping out interface regions that displayed task instructions or unrelated UI elements, and sampling one frame every five seconds to generate inputs for the VLM. This prevented instruction leakage and ensured the model inferred actions from observed behavior rather than visible text. The VLM then produced an ordered sequence of predicted user actions for each session, together with contextual cues derived from the spreadsheet interface.

To establish ground truth, we manually annotated whether each task in every video was actually performed by the participant. These annotations provided a binary reference for evaluation, enabling us to assess the VLM's predictions against observed behavior. 

\paragraph{Evaluation}
We evaluated the VLM's predictions against the annotated ground truth using accuracy, precision, recall, and F1 scores. Since each task in the dataset corresponds to an atomic spreadsheet operation (e.g., inserting a row, switching sheets, editing a formula), these metrics reflect the model's ability to recognize individual user actions rather than broader task success. To assess reliability of the ground truth, we also calculated the percent agreement and Gwet's AC1 to measure agreement between human annotations and an independent LLM (GPT-4.1) in labeling the VLM's outputs. Finally, to better understand the model's limitations, we analyzed error distributions across task types, identifying which spreadsheet operations the VLM most frequently failed to detect.

\subsection{Results}
\label{subsec:RQ1}

\paragraph{Recognition accuracy.}
We evaluated the VLM-based recognition pipeline by comparing its predicted action sequence against ground truth task lists collected for each session. Two independent raters---one human and one LLM-based---reviewed each VLM-identified action to determine whether it correctly matched a completed task in the session. The average F1 score was 90.5\% based on the human rater's judgments, and 90.4\% based on the LLM rater\footnote{Appendix~\ref{appendix:vlm-runtime} includes additional analysis of the VLM's processing time for these sessions.}.

\paragraph{Inter-rater agreement.}
To assess consistency between the two raters, we computed percent agreement and Gwet's AC1. The raters agreed on 96.4\% of action-level decisions ($\pm$ 5.4\%), with a Gwet's AC1 of 0.927 ($\pm$ 0.107), indicating strong agreement beyond chance.

\paragraph{Error analysis.}
Most recognition errors involved actions that occurred between sampled frames (e.g., rapid cell selections) or subtle UI changes (e.g., editing a formula in an adjacent cell). Confusion matrix analysis revealed that high-visibility actions like insertions and deletions were reliably detected, while visually subtle or transient actions were more error-prone. The most frequently missed actions were bold cell text, which was missed in 58.3\% of the sessions where participants performed this action; change font color (35.7\%), resize a column (18.8\%), create a new workbook (17.4\%), add a new row (13.0\%), and rename sheet (5.0\%).
\section{User Study}

To evaluate the effectiveness and user experience of \tool{}'s behavior-grounded recommendations, we conducted a within-subjects user study. Participants completed spreadsheet tasks using both \tool{} and a strong baseline: Excel Copilot, a prompt-based assistant purpose-built for spreadsheet support. Unlike general chat interfaces, Excel Copilot has direct access to spreadsheet content and is designed to generate context-specific recommendations based on user queries, making it a high-performing point of comparison.

Our study sought to evaluate whether \tool{}'s approach, observing user behavior through screen recordings and offering high-fidelity suggestions based on interpreted action traces, can help users discover and reflect on opportunities to improve their workflows. We organized our investigation around the following research questions:

\smallskip
\emph{RQ2. Can the system accurately identify inefficient or suboptimal workflows by observing user behavior alone?}

\smallskip
A core goal of our system is to detect inefficient workflows for improvement without relying on internal instrumentation (e.g., logs or APIs) or user prompts. We designed \tool{} to analyze screen recordings to provide targeted suggestions derived from semantic traces of user actions after completing the task (G1, G3). This question asks whether the system succeeds in identifying real inefficiencies in users' task execution based solely on visual observation.

\smallskip
\emph{RQ3. How do users interpret and respond to suggestions grounded in their actual behavior?}

\smallskip
We aim to generate suggestions that are not only accurate but also comprehensible and actionable. Suggestions should reflect what the user actually did, explain why the approach may be inefficient, and provide concrete alternatives (G2, G4). This question probes whether users find these behavior-grounded suggestions helpful, trustworthy, and conducive to reflection and learning.

\smallskip
\emph{RQ4. How does visual-grounded task reflection compare to prompt-based assistance in providing helpful guidance?}

\smallskip
In contrast to traditional AI assistants that wait for user input, \tool{} proactively identifies help opportunities by interpreting users' actions. This question explores how this proactive approach compares to the reactive model of prompt-driven assistance, particularly in terms of suggestion clarity, timing, and user preference.

\subsection{Methodology}

\subsubsection{Participants}
We recruited 20 participants through the Playbook UX platform. Participants spanned a wide range of occupations, including analysts (7), administrative staff (6), managers (4), accountant (1), engineers (1), and students (1). Their ages ranged from 23 to 56 years ($\mu = 34.4, \sigma = 9.3$). The sample included 12 participants who identified as male and 7 as female\footnote{One participant did not report gender.}.

Participants reported regular use of spreadsheet tools in their work: 75\% (15/20) used Excel or other spreadsheet software daily, while the remaining 25\% (5/20) used it weekly. In terms of experience, 50\% had more than 10 years of experience, 25\% had 6-10 years, and 25\% had 3-5 years.

We also asked about AI usage in spreadsheet tasks. While 45\% (9/20) of participants reported never using AI for spreadsheets, 35\% (7/20) used AI tools weekly, and 10\% (2/20) used them daily. The most commonly mentioned tools were ChatGPT and Copilot. Other tools included Gemini, Claude, and Excel's built-in Ideas feature. 

\subsubsection{Procedure}
Each participant completed a one-hour study session. After providing informed consent, participants first filled out a brief background questionnaire about their prior experience with Excel. To minimize potential bias, we referred to tools using neutral names (``Tool A'' for \tool{} and ``Tool B'' for Excel Copilot) and counterbalanced the order in which participants used them. The session proceeded through three phases: a data challenge task, a tutorial for both tools, and an evaluation session where participants reviewed and assessed recommendations generated by both tools. We chose not to show suggestions during the tasks to avoid disrupting participants' focus or altering their natural workflows. Instead, suggestions were shown only after task completion to support reflective learning, aligning with our goal of providing just-after-it-mattered guidance (G4).

\paragraph{Data Challenge Task.}
Participants completed one of two realistic Excel-based data challenges designed to simulate multi-step workflows in large, multi-sheet workbooks. The design of these challenges was inspired by Kaggle-style analytics problems, where participants work with realistic datasets to answer open-ended analysis questions. Each challenge contained four questions involving tasks such as summarizing trends, analyzing category-based performance, and identifying opportunities for improvement (see Appendix~\ref{appendix:tasks} for task details). To maximize coverage of different spreadsheet operations, each participant completed only one of the two challenges, and task assignment was counterbalanced across participants. Pilot testing with five participants showed that the two challenges were matched in both completion time and perceived complexity. Each question in the challenge could be completed using either simple but repetitive operations (e.g., filtering, sorting, and basic formulas) or more advanced features such as \texttt{SUMIF} formulas and PivotTables, enabling us to evaluate system effectiveness across diverse workflows and experience levels. During the task, the experimenter recorded the entire window shared by the participants for later analysis.

\paragraph{Tutorial.}
Before the suggestion evaluation, participants were given five minutes to familiarize themselves with each tool through a short tutorial. In the baseline condition, participants typed their descriptions of recent steps and context into a text input box. In contrast, with \tool{}, participants clicked the ``Show me suggestions to improve my workflow'' button (Figure~\ref{fig:interface}, \textcolor[HTML]{595959}{\ding{202}}) to receive the first suggestion, and could then use the ``Give me another suggestion'' option (Figure~\ref{fig:interface}, \textcolor[HTML]{595959}{\ding{206}}) to request additional suggestions (three suggestions in total). The features introduced during this tutorial were identical to those available during the evaluation session.

\paragraph{Suggestion Evaluation Session.}
After completing the data challenge, participants entered a recommendation evaluation session where they used both tools to retrieve, review, and optionally test three recommendations per tool. The order of tools was counterbalanced. In the baseline condition, participants were instructed to prompt the tool to generate suggestions. 
In the \tool{} condition, recommendations were generated automatically from participants' screen recordings and displayed by the experimenter in the Excel interface. Screen recording was initiated by the experimenter at the start of the task to support VLM processing, but participants were informed that in a fully integrated version of \tool{}, this would occur automatically.
For each suggestion, participants could decide whether to test it on the same spreadsheet used in the earlier task. We logged whether they attempted the suggestion, and whether the suggested steps, when followed, produced the correct outcome.

\paragraph{Questionnaires.}
Participants completed three sets of questionnaires.\footnote{Full questionnaires are provided in the supplemental materials.} At the start, a brief background questionnaire collected participants' prior experience with Excel. During the recommendation evaluation session, after reviewing each suggestion, participants rated it on eight 5-point Likert-scale items capturing multiple aspects of the suggestion (e.g., usefulness, clarity, relevance, learning value; see supplemental materials for full list). At the end of the session, a final questionnaire captured participants' overall impressions, including perceived workload, effort required to obtain suggestions, privacy concerns, and preferences between tools.

\subsubsection{Measurements and Analysis}
After each suggestion, participants provided 5-point Likert ratings on eight dimensions. We analyzed these ratings using linear mixed-effects models, with fixed effects for tool, task, tool order, whether the participant attempted to apply the suggestion, whether the suggestion produced the correct result when tested, and the number of distinct tools referenced in the suggestion. A random intercept was included for participant ID. Significance was assessed using F-tests with Satterthwaite's approximation of degrees of freedom~\cite{satterthwaite1946approximate}.

For the three summary-level questions on overall impressions, perceived privacy confidence, effort required to obtain help, and system preference, we used two-tailed Wilcoxon signed-rank tests~\cite{wilcoxon1992individual}. All $p$-values were adjusted using the Holm-Bonferroni method~\cite{holm1979simple}, with a significance threshold of $\alpha = 0.05$.

We also conducted a thematic analysis~\cite[Chapter 5]{blandford2016qualitative} of participants' open-ended responses to identify recurring themes around perceived strengths, limitations, and ideal usage scenarios. 

\begin{figure}
    \centering
    \includegraphics[width=\linewidth]{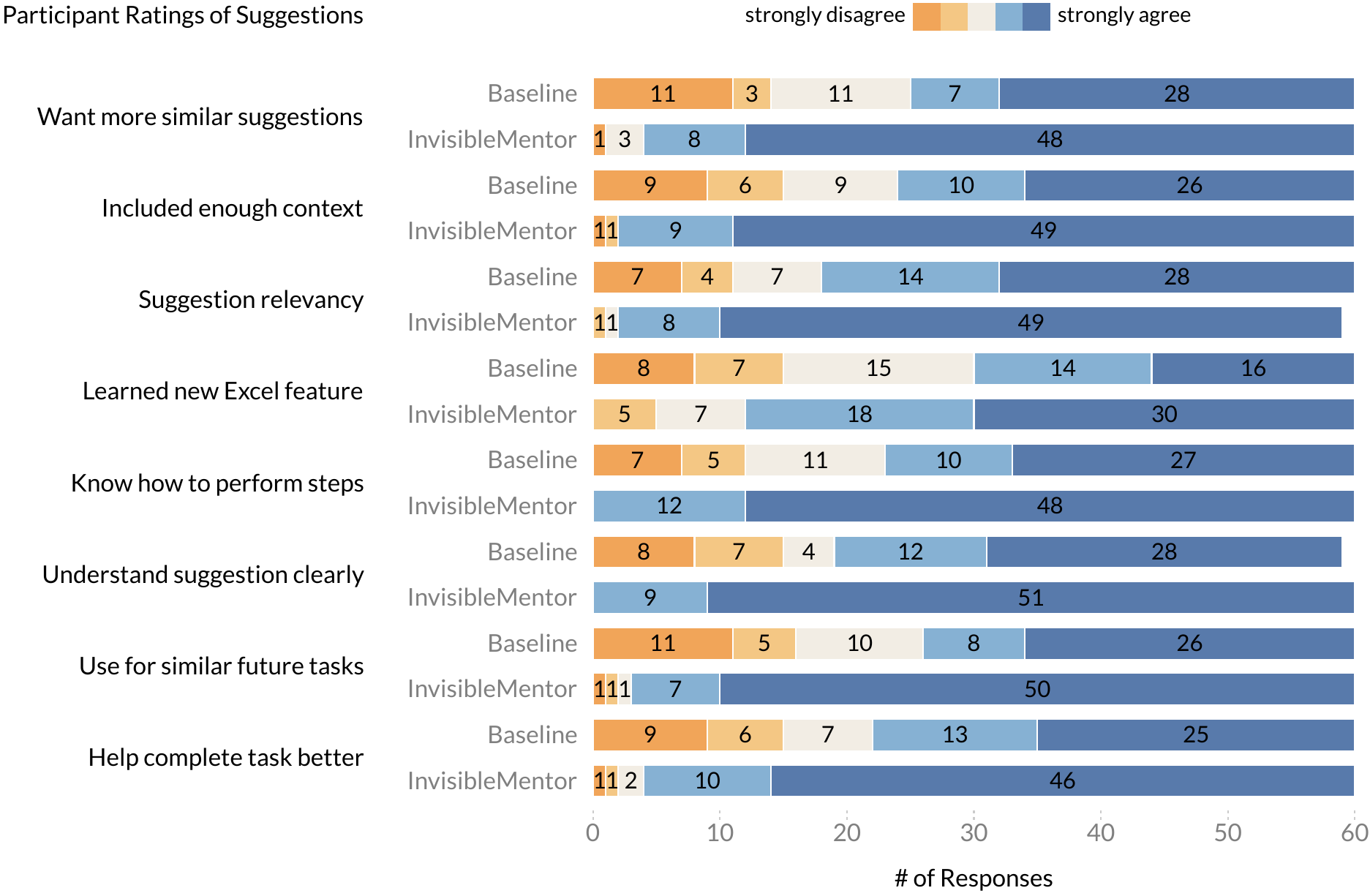}
    \vspace{-2em}
    \caption{Participant ratings of \tool{}'s suggestions. \textmd{Stacked bar charts show agreement levels with eight evaluative statements on a 5-point Likert scale (from ``Strongly Disagree'' to ``Strongly Agree''). Across all statements, participants rated \tool{}'s suggestions as significantly more useful, understandable, and better aligned with their recent tasks than those from the baseline system.}}
    \label{fig:suggestion_ratings}
    \Description[]{}
\end{figure}

\subsection{Results}

\subsubsection{RQ2: \tool{} identifies real inefficiencies}
\label{subsec:RQ2}

\paragraph{\tool{} detected inefficient or error-prone workflows from screen activity}
Most participants ($P1$, $P2$, $P7$–$8$, $P11$–$14$, $P20$) described \tool{}'s suggestions as tightly linked to their own actions, contrasting it with the baseline's responses. They found that \tool{} knew ``what I was trying to do'' ($P8$), ``tracked what I performed'' ($P13$), and ``highlighted the things I did manually'' ($P2$). $P1$ noted that \tool{} ``went based off of my attempts and edited it as I was attempting it,'' while $P11$ suggested enabling mid-task delivery to make suggestions even more timely.

\tool{} uncovered several commonly recurring inefficiencies across participants' workflows, including: repeatedly using Find and Replace to standardize inconsistent categorical values like gender ($P2$, $P11$, $P19$), manually writing separate SUM formulas for each month instead of using SUMIF or PivotTables ($P1$, $P5$, $P11$, $P20$), and frequently copying tables between spreadsheets to apply formulas like VLOOKUP ($P12$, $P15$, $P18$).

\subsubsection{RQ3: Users find behavior-grounded suggestions comprehensible and actionable}
\label{subsec:RQ3}

\paragraph{\tool{} provided clear, easy to follow, and confidently actionable suggestions}
Participants found \tool{}'s suggestions significantly easier to understand than those of the baseline ($F = 25.1$, $p < 0.001$)\footnote{One participant did not provide a rating for the item ``understand suggestion clearly'' on one of the suggestions.}, and reported greater confidence in knowing how to carry out the recommended actions ($F = 29.3$, $p < 0.001$). This perceived clarity was reflected in participants' comments: $P11$ described the suggestions as ``short, concise, and straight to the point,'' while $P5$ noted they were ``easier to follow.'' In contrast, $P10$ found the baseline ``too complex to execute,'' and $P20$ remarked they ``didn't know if [baseline suggestions] were true or not.''

\paragraph{\tool{}'s suggestions reflected the user's recent actions and intent}
Participants rated \tool{}'s suggestions as significantly more relevant to their just-completed tasks ($F = 15.7$, $p < 0.001$)\footnote{One participant did not provide a rating for the item ``suggestion relevancy'' on one of the suggestions.}, and more likely to teach them something new about Excel ($F = 11.6$, $p = 0.01$). They also reported receiving more contextual explanations for why a suggestion was offered ($F = 27.6$, $p < 0.001$). $P1$ said that the tool ``gave me corrections based on my past attempts,'' while $P8$ noted that it was ``actively guessing what I'm trying to do to make suggestion.''

\begin{figure}
    \centering
    \includegraphics[width=\linewidth]{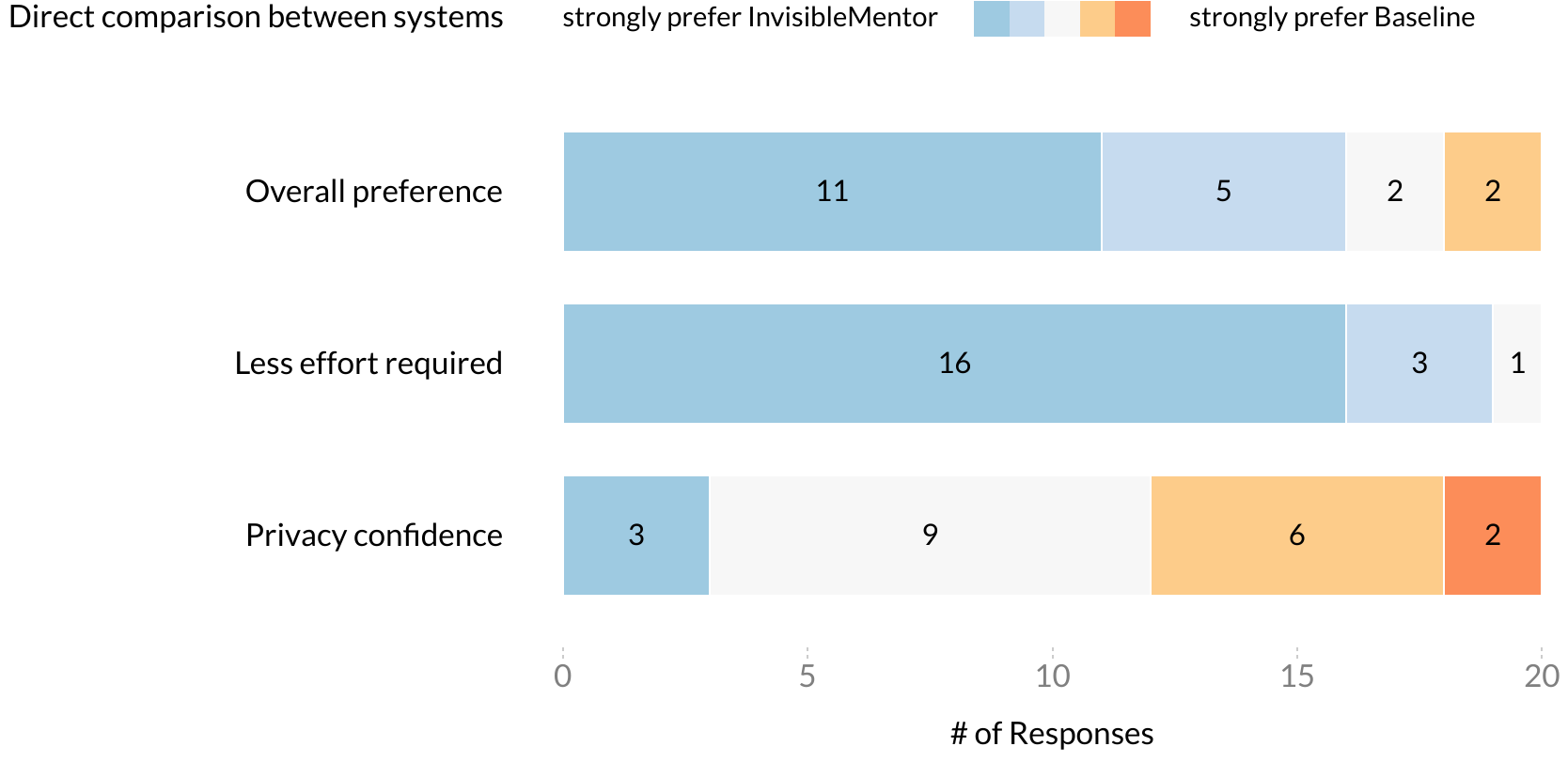}
    \vspace{-1em}
    \caption{Participants' comparative preferences between \tool{} and the baseline. \textmd{Participants were asked which tool required less effort, which they would overall more prefer to use in the future, and which they trusted more with regard to privacy. Most participants significantly preferred \tool{} for effort and rated it as overall more preferable to use, with no significant difference in perceived privacy confidence.}}
    \label{fig:forced_choice}
    \Description[]{}
\end{figure}

\subsubsection{RQ4: Proactive behavior-based suggestions reduce effort and support learning}
\label{subsec:RQ4}

\paragraph{\tool{} reduced users' effort compared to prompt-based alternatives}
Participants rated \tool{} as requiring significantly less effort to use than the baseline ($W = 0.0$, $p < 0.001$), and expressed an overall preference for it ($W = 8.0$, $p < 0.001$). Despite the system's use of screen recordings to track user behavior, participants did not report heightened privacy concerns ($W = 27.0$, $p = 0.58$) compared with baseline.

\paragraph{\tool{}'s suggestions supported learning and workflow reflection}
Participants rated \tool{}'s suggestions as significantly more useful for improving their workflows than those from the baseline ($F = 18.8$, $p < 0.001$). They reported greater willingness to reuse \tool{} in real-world settings ($F = 23.7$, $p < 0.001$), and expressed interest in receiving more similar suggestions in future work ($F = 18.6$, $p < 0.001$) than baseline. Several participants described the tool as acting like a tutor ($P5$, $P13$, $P15$, $P17$, $P19$), offering context-sensitive feedback that promoted reflection and growth ($P1$–$2$, $P8$, $P11$, $P13$, $P15$, $P17$, $P19$). $P5$ remarked that the tool ``taught me more and quicker,'' and $P19$ said it helped them ``learn and apply these corrections.''

In contrast, some participants described the baseline as limited in supporting reflection or deeper learning ($P3$, $P5$-$6$, $P8$-$10$, $P17$, $P19$). $P10$ remarked that while the baseline ``felt more advanced,'' its suggestions were ``too complex to execute,'' limiting their educational value. $P6$ observed that the baseline ``would give you a suggestion dependent on what you ask.'' $P17$ noted that the baseline was ``better for getting work out there quickly,'' but less useful for learning and development. While most participants saw limited learning value in the baseline, some acknowledged occasional benefits ($P4$, $P8$, $P14$, $P18$). $P4$ shared that the baseline ``allowed me to personalize the feedback and support,'' and appreciated that both systems ``highlighted things I wouldn't have initially thought about.'' $P18$ similarly described the baseline as helpful for ``a quick answer and fix.'' 

They also viewed baseline suggestions were often perceived as one-off fixes ($P9$, $P17$-$19$). $P9$ noted that while baseline ``would help me now,'' it was \tool{} that ``would help me for the future.'' $P19$ similarly reflected that baseline ``had a tendency to try and do things for me,'' which ``may be helpful in the moment but maybe I would be less likely to apply the suggestions in the future.''

\section{Discussion}
\label{sec:discussion}
\subsection{Revisiting Our Study Questions}
\tool{} accurately recovered user actions from screen recordings using a VLM, demonstrating strong agreement with ground truth annotations (Section~\ref{subsec:RQ1}, RQ1). By analyzing users' observable behavior, it discovered inefficient or repetitive workflows such as repeated Find \& Replace operations or redundant manual summaries (Section~\ref{subsec:RQ2}, RQ2). Participants described structured, high-fidelity suggestions as easier to understand and more contextually appropriate than baseline assistance, and reported that they supported learning and workflow improvement (Section~\ref{subsec:RQ3}, RQ3). Compared to prompt-based suggestions, participants rated \tool{}'s output as more useful, actionable, and reflective of what they had just done, expressing stronger interest in future use (Section~\ref{subsec:RQ4}, RQ4).

\paragraph{Other factors influencing suggestion ratings.}
To understand what else might shape participants' ratings, we analyzed several additional factors, including the task being performed, tool usage order, the number of distinct Excel features referenced in a suggestion, whether participants chose to test the suggestions during the study, and whether the suggestions, when tested, successfully produced the intended outcome. Among these factors, we found no significant effect of task or tool usage order on suggestion ratings across any of the evaluation questions ($p > .1$ and $p > .3$, respectively).

However, the number of distinct Excel features referenced in a suggestion showed significant effects on several measures, including how clearly participants understood the suggestion ($F = 11.0$, $p = 0.006$, $\beta = 0.14$), whether they learned new Excel features ($F = 8.7$, $p = 0.019$, $\beta = 0.14$), and whether the suggestion included enough context ($F = 9.9$, $p = 0.011$, $\beta = 0.14$). These results show that suggestions referencing a wider range of built-in Excel features were perceived as clearer, more educational, and better contextualized. Participants who tested suggestions, or for whom the suggestions successfully produced the correct outcome, also gave significantly higher ratings of relevance ($F = 6.5$, $p = 0.049$ and $F = 8.3$, $p = 0.024$, respectively) and were more willing to receive similar suggestions in the future ($F = 6.8$, $p = 0.042$ and $F = 9.0$, $p = 0.017$).

\subsection{Revisiting Our Goals}

\subsubsection{G1. Translate visual representations into semantically rich representations without specialized instrumentation.}

Rather than relying on instrumentation or logs, \tool{} interprets workflows directly from visual evidence. This grounding in actual behavior led participants to describe the system's suggestions as specific, trustworthy, and aligned with their needs. $P3$ said it ``felt like the system had been watching me and knew exactly where I messed up.'' $P6$ appreciated how it ``pointed out my error and mistypes,'' and $P13$ valued that it identified ``where there may be data breakage and how to solve for them.'' Participants found that the suggestions reflected not just generic advice, but a structured understanding of what they had done, contributing to higher relevance and usefulness.

While Excel provides built-in mechanisms such as VBA macros to log user actions, these methods require users to enable macros and are limited to capturing interactions within a single spreadsheet. Our goal was not to replace such mechanisms, but to explore an alternative that is more lightweight, setup-free, and broadly applicable. By relying solely on screen recordings, our system avoids instrumentation barriers and can reconstruct workflows that span multiple files and applications. This makes \tool{} particularly useful for generating suggestions or feedback after a task is completed, even when no prior logging or setup was in place.

\subsubsection{G2. Provide structured, high-fidelity suggestions grounded in user behavior.}

\paragraph{\tool{}'s suggestions included step-by-step explanations and examples}
Beyond clarity, participants valued that \tool{} provided concrete, example-driven guidance. Several participants ($P3$–$4$, $P10$, $P12$, $P14$, $P17$, $P20$) mentioned that the tool didn't merely point out issues, but offered detailed next steps. $P12$ highlighted the system's ``step by step examples to walk you through,'' and $P20$ similarly appreciated the ``step-by-step method and explanation of the formulas.'' $P4$ emphasized the importance of ``list[ing] out in detail what I need to do,'' while $P17$ praised the ``clear steps and context.''

\paragraph{\tool{}'s suggestions aligned with participants' skill levels and data contexts}
Participants ($P3$, $P5$–$6$, $P10$-$11$, $P15$, $P17$) reported that the suggestions met them at the right level of detail and complexity. $P5$ explained that the guidance was ``shorter and more concise,'' helping them move faster under time pressure. $P6$ noted that they could ``build upon [the suggestion],'' and $P17$ appreciated that the system adapted to larger datasets and included justification for why one method was superior.

\subsubsection{G3. Discover unspoken help opportunities by modeling user behavior directly from observation.}

Most participants ($P1$-$3$, $P5$-$9$, $P11$, $P13$, $P15$, $P19$) attributed \tool{}'s low-effort interaction to its ability to infer intent from user actions, which eliminated the need to craft explicit prompts. $P3$ noted that the tool ``felt quicker to prompt based on what I was already doing without having to directly posit a question,'' emphasizing the reduced need for explicit formulation. $P1$ appreciated that \tool{} ``edited [suggestions] as I was attempting it,'' highlighting its ability to offer in-the-moment guidance without user intervention. $P7$ further remarked that \tool{} ``interpreted your actions so it knew where to prompt you... and highlighted observing actions, which is something I have not seen in other AIs,'' while contrasting this with the baseline, which ``was just asking raw questions based on what you thought.'' Furthermore, $P8$ described the baseline as ``more like a how-to-do guide'' that required users ``to be more specific to use [it] effectively.''

\paragraph{Behavior-based analysis revealed overlooked opportunities}
Participants also noted that \tool{} identified workflow improvements they might not have noticed on their own. For example, $P2$ remarked that the tool highlighted things ``that I did manually which can break or mess up the data.'' $P19$ described how \tool{} ``offered better solutions for the future,'' and $P13$ shared that the system ``highlighted other ways to accomplish the task... that I did not think of.'' 

\subsubsection{G4. Deliver guidance shortly after task completion to promote reflective learning.}

\paragraph{Participants saw long-term learning value in post-task suggestions}  
Many participants ($P3$, $P5$, $P9$-$P10$, $P13$, $P15$, $P17$-$P19$) appreciated that \tool{} delivered suggestions after they had completed the task, enabling them to reflect on their actions and consolidate learning. $P15$ noted that the system helped them ``understand how to do things better,'' and $P17$ likened it to a study guide that ``scales with larger datasets,'' emphasizing its support for recurring tasks. By receiving feedback after completing the task, participants could take time to review missteps and internalize new strategies without interrupting their workflow. $P2$ shared that they would reuse the tool ``to pinpoint errors and work from there,'' indicating a preference for reflection over immediate correction.

There are a few participants ($P2$, $P11$–$12$, $P20$) who desired real-time suggestions. $P11$ explicitly asked whether \tool{} could provide help ``in the middle of the task,'' and $P12$ said ``I do find the value of [\tool{}] if I can get more in-real time suggestions''

\subsection{Limitations}
\label{sec:limitations}

Our findings should be interpreted in light of several limitations. First, our evaluation focused on a set of spreadsheet tasks that were intentionally chosen to be common and well-scoped (e.g., sorting, filtering, and simple formula edits). While these tasks represent realistic use cases, they may not reflect the full diversity of spreadsheet workflows. For example, tasks involving complex macros, cross-sheet dependencies, or domain-specific logic (e.g., financial modeling) may pose challenges for the current system's suggestion quality and reliability.

Second, \tool{} relies heavily on accurate recognition of user actions from screen recordings. Although our evaluation showed high accuracy for the chosen task types, some actions are harder to detect or interpret. For example, bolding one cell among many selected cells can be visually subtle, and switching between formula view and normal view may involve only minor interface changes, making them more ambiguous for the system to recognize. As shown in Section~\ref{subsec:RQ1}, brief or transient actions such as changing font color or resizing a column are especially prone to omission due to the VLM's frame selection mechanism, which may skip over these short intervals. Such misrecognitions can reduce the relevance of generated suggestions or introduce hallucinated explanations that do not reflect the user's actual behavior.

Third, our study participants were a mix of novice and spreadsheet-proficient users. While most had prior experience with Excel and AI tools, their behaviors may differ from those of domain experts (e.g., accountants, analysts) or novices. Additionally, the study was conducted in a controlled lab setting. In real-world workflows, time pressure, multitasking, or incomplete task context may affect how users perceive or act on system suggestions.

Finally, while participants generally reacted positively to the idea of AI-generated suggestions, their actual reliance on them was often opportunistic and selective. Some participants ignored suggestions they found too obvious or irrelevant, while others expressed hesitation to fully trust AI recommendations. These nuanced behaviors point to the importance of ongoing refinement in timing, relevance, and trust calibration for such systems.

\subsection{Future Work}

\paragraph{Scaling to more diverse and unconstrained workflows}
To extend beyond controlled settings, future work can further evaluate our system on large-scale real-world recordings, such as Excel tutorial videos from YouTube or screen-captured workplace tasks. These sources offer access to a broad spectrum of workflows, including noisy and unstructured environments where the user's goals are more implicit and actions less cleanly segmented. By analyzing such recordings, the system could identify inefficient patterns (e.g., manual recalculations, redundant copying) and underused features (e.g., filters, pivot tables), uncovering new opportunities for automation or guidance. Combining visual inputs with lightweight telemetry, such as formula edit logs, cursor dwell time, or tab focus events, could improve robustness in noisy conditions and help better infer task boundaries. This would also help move beyond per-frame analysis to more semantically coherent action modeling.

\paragraph{Strengthening reliability with self-correction and feedback loops}
Our current system offers no way to revise or verify misinterpreted actions, which can propagate errors into the generated suggestions. Future iterations could incorporate self-correction strategies that detect low-confidence predictions and re-query recent frames, or allow users to ``backtrack'' and adjust a misunderstanding. For instance, agents might flag uncertain interpretations with a soft highlight and ask the user for confirmation (``Did you mean to insert a row here?''). Lightweight verification via user ratings or in-context edits could be used to fine-tune the system online, without requiring full labeling. A key challenge is balancing helpfulness with intrusiveness: too many interruptions could overwhelm users, while too few could undermine trust.

\paragraph{Generalizing the framework to other domains}
While our evaluation focused on spreadsheets, our framework, extracting visual actions from screen recordings and generating structured, high-fidelity suggestions, can extend to other domains. One promising direction is interactive visualization dashboards (e.g., Tableau, Power BI), where users interactively explore data by filtering, encoding, and transforming views.

Our design goals (G1–G4) offer a lens for thinking about this transfer. G1 remains broadly applicable. Like Excel, dashboards expose rich visual states (e.g., axis configurations, filter panels, view changes) that can be tracked from screen recordings without requiring instrumentation or backend access. G2 would still emphasize concrete feedback (e.g., ``You toggled between pie charts and bar charts repeatedly---consider using a grouped bar chart for clearer comparisons''). However, suggestion logic must evolve to accommodate exploratory workflows, which often lack a fixed endpoint. Rather than flagging violations of best practices, suggestions could surface alternative strategies or reminders of missed comparisons based on the user's exploration path. G3 is especially valuable in dashboards, where users often fail to notice biases (e.g., only slicing by a single category) or overlook available chart types. Our visual observation approach could proactively detect such moments without waiting for user prompts. G4 would need adaptation: dashboard tasks are often nonlinear and ongoing. Still, post-task summaries (e.g., ``You filtered by time range but did not compare revenue across segments'') could aid users in reflecting on their analysis coverage or gaps, especially when used repeatedly across sessions.

While adapting our system to this domain would require new domain-specific primitives (e.g., mark types, view compositions), our broader philosophy, providing suggestions in observed visual activity, remains powerful in helping users learn from their own behavior.
\section{Conclusion}

We presented \tool{}, a system that generates high-fidelity, post-task suggestions grounded in users' screen-recorded behavior. Unlike prior assistants that rely on logs, APIs, or user prompts, \tool{} observes visual actions and infers improvement opportunities through a two-phase pipeline of action reconstruction and suggestion generation. Our evaluation shows that \tool{} not only recognizes inefficient workflows with high accuracy but also supports reflection and future learning through structured, behavior-specific guidance. These findings demonstrate the potential of visual activity–based assistance in promoting more efficient, learnable workflows, without requiring internal instrumentation or user query formulation.

\begin{acks}
\end{acks}

\bibliographystyle{ACM-Reference-Format}
\bibliography{references}
\balance{}
\appendix
\section{Prompt Templates}
\label{sec:appendix-prompts}

This appendix provides the prompt templates used in each phase of \tool{}'s architecture. The prompts were constructed to guide the vision-language and language models in extracting meaningful task representations and generating grounded, actionable recommendations. These prompts were issued via the OpenAI API during system execution.

\subsection{VLM Prompt for Task Representation Extraction}

We use the GPT-4.1 vision-language model to extract structured user actions and spreadsheet content from screen recordings. The prompt is shown below:

\begin{lstlisting}
You are an assistant for workflow analysis. Given a sequence of frames from a task video and a list of prior identified actions, analyze the frames and identify any new user actions that are not already described in the prior actions. If you find new actions, add them to the new_actions. If no new actions are identified, return empty new_actions.

You will be provided with a batch of frames and a list of prior actions. For each action, you need to identify all of the details, such as the formatting, "cell content", "formula", and the specific action. Notice that some actions may have different outcomes based on the content of the sheet. For example, if the user switches to a different sheet, it may create a new sheet in addition to switching to it. 

For some actions not shown on the frames, you need to predict them by comparing the difference between the frames. For example, whether the styles, formatting, or content of the cells have changed. If the user has changed the content of a cell, you need to speculate the possible actions that may lead to the changes and add the actions to the new_actions.

If the content of the sheet has changed (not cell formats), you need to record the entire sheet in Markdown format. Remember to include the workbook name and sheet name.

Your response should be a JSON object with the following structure:
{   
    "new_action_detected": true/false,
    "new_actions": [
        "action1",
        "action2",
        ...
    ],
    "sheet_changes": true/false,
    "sheet_details": "use Markdown format to record the entire sheet"
}
\end{lstlisting}

\subsection{LLM Prompt for Recommendation Generation}

To generate recommendations, we use a textual LLM (OpenAI o3 model) prompted with a structured representation of the user's observed actions and spreadsheet state. The prompt instructs the model to identify suboptimal workflows, explain their inefficiencies, and suggest improved methods using built-in Excel features. The model is asked to provide step-by-step instructions grounded in the user's context, and to rank recommendations by impact.

\begin{lstlisting}
You are a workflow efficiency expert. Analyze user actions from Excel task videos and identify suboptimal workflows.

Instructions:
1. Group related actions into workflows (steps accomplishing a specific task)
2. For each workflow, set "Optimal" to true/false based on efficiency
3. For suboptimal workflows ("Optimal": false):
   - "ActionList": List actions starting with "It looks like you..."
   - "Reason": Main inefficiency (be specific) starting with "You ..."
   - "Suggestion": Provide ONE actionable solution using Excel features:
     - Give step-by-step instructions with exact Ribbon paths/shortcuts
     - Include detailed examples with realistic sheet/column names
     - Prioritize automation over manual repetition
     - Provide complete formulas with explanations when applicable
     - End with "Benefit:" explaining concrete improvements (time saved, fewer steps, error reduction)
     - Compare before/after: "Original: X steps, Suggested: Y steps"

4. Focus on efficiency and maintainability, not just task completion
5. Only include 3 most impactful suboptimal workflows and rank them by importance
6. Use proper formatting: backticks (`) around Excel functions, formulas, keyboard shortcuts, and feature names, and triple backticks (```) for multi-line formulas or step-by-step code examples
7. Create plausible placeholders for unclear data references

Output JSON format:
{
    "Workflows": [
        {
            "ActionList": ["Action 1", "Action 2"],
            "Optimal": true/false,
            "Reason": "Brief explanation",
            "Suggestion": "Step-by-step actionable solution"
        }
    ]
}
\end{lstlisting}

\section{User Scenario}
\label{sec:scenario}

\newcommand{\persona}{Mia}
\newcommand{\Pronoun}{She}
\newcommand{\pronoun}{she}
\newcommand{\Possessive}{Her}
\newcommand{\possessive}{her}

\persona{}, an analyst working with sales data, wants to calculate the total revenue for each city. Her goal seems straightforward, but she isn't sure of the best approach. She tries several methods: filtering rows manually, using combinations of \texttt{SUM()} and \texttt{IF()} formulas, hovering over columns to check auto-totals, and eventually constructing a \texttt{UNIQUE} column with corresponding \texttt{SUMIF()} formulas. These attempts span different sheets, ranges, and parameter values, many of which she adjusts or discards along the way.

Later, she wonders if there's a more efficient solution, and considers asking a chatbot. But expressing her workflow turns out to be harder than expected. She vaguely recalls that \texttt{SUMIF} ``didn't work'' at first, but doesn't remember why. It may have been due to a mismatched range, but she can't verify it. In describing her steps, she omits earlier failed attempts and small but relevant actions like adjusting cell formats or deleting helper columns. Even when she tries to be thorough, she struggles to name certain tools (``that dropdown with formula functions'') or describe all the cell styles she modified. Faced with the burden of reconstructing everything in words, she gives up on asking for help.

Instead, she turns to \tool{}, which generates recommendations based on her observed interactions, without requiring her to describe them (G1). In the sidebar, one of the existing five prompt ideas has been repurposed to serve as an entry point for \tool{}, and \persona{} selects it (Figure~\ref{fig:interface}, \textcolor[HTML]{595959}{\ding{202}}). This selection issues a pre-populated request on her behalf, triggering the assistant to display a recommendation grounded in her actual workflow.

Next, \tool{} summarizes Mia's recent activity, such as building a summary table with \texttt{UNIQUE} and \texttt{SUMIF()} formulas, and highlights limitations of this approach (Figure~\ref{fig:interface}, \textcolor[HTML]{595959}{\ding{203}}, \textcolor[HTML]{595959}{\ding{204}}). By tying its suggestions to specific steps in her process, the system ensures its feedback is directly relevant.

To help her improve, \tool{} offers a clear, step-by-step guide for using PivotTables to achieve the same goal more efficiently (Figure~\ref{fig:interface}, \textcolor[HTML]{595959}{\ding{205}}) (G2). This recommendation appears immediately after she finishes building her summary table, while the task is still salient and her cognitive load has eased (G4). She can also request alternative suggestions using a dedicated ``give me another suggestion'' option (Figure~\ref{fig:interface}, \textcolor[HTML]{595959}{\ding{206}}).

With minimal effort, \persona{} produces a more maintainable summary and learns a new feature that improves her workflow. The assistant bridges the gap between what she did and what she could have done better, without requiring her to reconstruct her process or translate it into a prompt (G3).

\section{Additional Details for Evaluation 1}
\subsection{Task Overview and Spreadsheet Contents}
\label{appendix:eval1-details}
In Evaluation 1, we used a set of benchmark tasks that span a range of common spreadsheet activities. These tasks were performed in the browser using a standardized Excel file containing two sheets:

\begin{itemize}
  \item \textbf{Budget sheet}: Includes 7 rows (Title, empty row, and 5 categories---Category, Rent, Car, Food, Total) and 5 columns (Category, Jan, Feb, Mar, Apr). Cells were pre-filled with monthly budget data (e.g., Rent = \$1,275).
  \item \textbf{Address History sheet}: Contains 3 rows of addresses with columns for Street Line 1, Line 2, City, State, Zip Code, and Move Date (From / To). Example: `33 Cherry Drive, N/A, San Diego, CA, 92104, June 2021, Present`.
\end{itemize}

Participants completed the following 14 tasks using the Excel web app:

\begin{enumerate}
  \item \textbf{Open file}: Access a spreadsheet via a link in a Gmail message.
  \item \textbf{Add row}: Insert a new row (e.g., a new category) in the Budget sheet.
  \item \textbf{Add values}: Input numerical values (e.g., monthly expenses) into the spreadsheet.
  \item \textbf{Edit formula}: Modify or insert a SUM formula to update a monthly total.
  \item \textbf{Bold}: Apply bold formatting to cell text.
  \item \textbf{Cell fill color}: Use fill color to highlight a cell (e.g., coloring column header).
  \item \textbf{Switch sheet}: Navigate between the `Budget` and `Address History` sheets.
  \item \textbf{Resize column}: Adjust the width of a column (e.g., Street Name column).
  \item \textbf{Copy/paste}: Duplicate content within the spreadsheet.
  \item \textbf{Change font color}: Modify the font color of selected cells.
  \item \textbf{New sheet}: Create a new, blank worksheet in the workbook.
  \item \textbf{Rename sheet}: Rename an existing worksheet (e.g., rename `Sheet1` to `Accounts`).
  \item \textbf{Share link}: Generate and share a link to the workbook via email.
  \item \textbf{New workbook}: Start a new workbook from the Excel start screen.
\end{enumerate}

These tasks were selected to cover a range of editing, formatting, navigation, and collaboration operations commonly encountered in everyday spreadsheet use.

\subsection{VLM Processing Time and Efficiency}
\label{appendix:vlm-runtime}
\begin{figure}
    \centering
    \includegraphics[width=\linewidth]{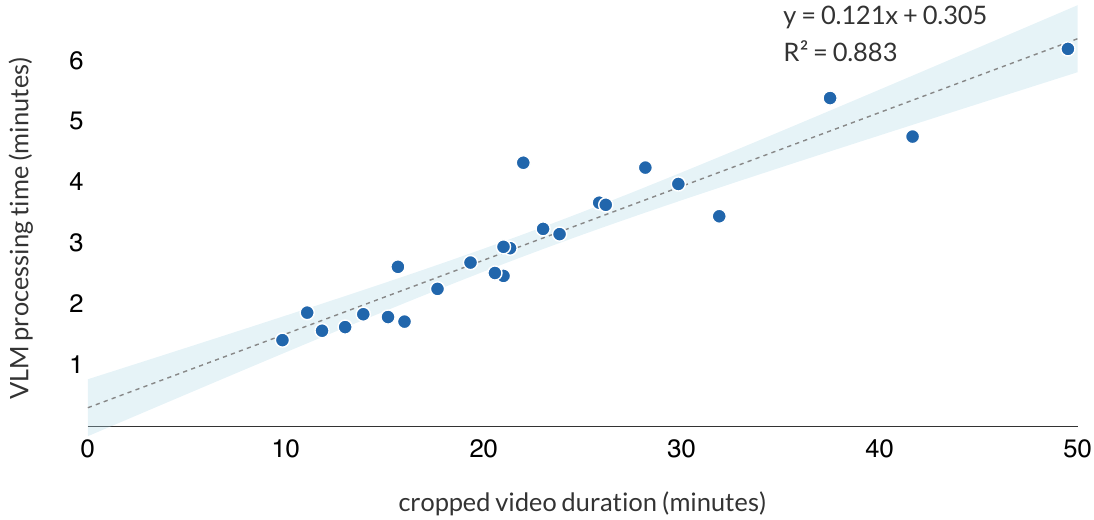}
    \caption{Relationship between video duration and VLM processing time. \textmd{Each dot represents one screen recording session from Evaluation 1. The x-axis represents the duration of the cropped input video, and the y-axis shows the time taken by the VLM to process that session. A fitted linear regression line indicates a strong positive correlation ($R^2 = 0.883$).}}
    \label{fig:processing-time}
    \Description[]{}
\end{figure}

In addition to measuring action extraction accuracy, we recorded how long the Vision-Language Model (VLM) took to process each benchmark screen recording during Evaluation~1 (Figure~\ref{appendix:vlm-runtime}). Because participants took varying amounts of time to complete each task, the video durations ranged from 9.8 minutes to 49.5 minutes. 

We measured the wall-clock time required for the VLM to process each video, recorded during Evaluation~1 under the same experimental setup (see Section\ref{sec:eval1-setup}). Frames were sampled at fixed 5-second intervals and passed to the model for inference. The average processing time was 2.99 minutes ($SD = 1.18$). The shortest video (9.8 minutes) took 1.41 minutes, while the longest (49.5 minutes) took 6.20 minutes. Figure~\ref{fig:processing-time} shows a dot plot of video duration versus processing time, with a fitted regression line indicating a strong linear relationship.

\section{Task Instructions}
\label{appendix:tasks}

This appendix provides the full task descriptions shown to participants during Study 2. Tasks were designed to simulate realistic spreadsheet analysis scenarios involving multi-step operations over large, multi-sheet workbooks.

\subsection{Task A: Vrinda Store Sales Analysis}
Vrinda is an exclusive clothing store offering a wide range of men's and women's wear tailored to various age groups. The goal of this task is to analyze Vrinda's sales data for the year 2022 to gain insight into the customer base and identify opportunities for revenue growth.

\paragraph{Task Description.} Using Microsoft Excel and the provided dataset, please complete the following analysis tasks:
\begin{enumerate}
    \item Create a single chart that compares total sales (Amount) and the number of orders (Qty) by month for 2022.
    \item Identify which month had the highest total sales and the most orders.
    \item Analyze whether women or men made more purchases in 2022.
    \item Use the 2022 sales trend to forecast monthly sales for 2023.
\end{enumerate}

Please spend about 8 minutes on each question and do your best within that time. Perfection is not expected. You will be notified when one minute remains, and when it is time to move on to the next task. There are no right or wrong answers---we are simply interested in seeing your natural approach.

\subsection{Task B: Pizza Sales Analysis}
In this task, you take on the role of a BI consultant hired by Plato's Pizza, a Greek-inspired restaurant in New Jersey. Your objective is to analyze the restaurant's performance using historical order data and identify operational insights.

\paragraph{Task Description.} Using Microsoft Excel and the provided dataset, please complete the following analysis tasks:
\begin{enumerate}
    \item Determine which days of the week and which times of day tend to be the busiest.
    \item Estimate how many pizzas are made during peak periods.
    \item Identify the best-selling and worst-selling pizzas.
    \item Calculate the average order value.
\end{enumerate}

As before, spend approximately 8 minutes on each question. You will be notified near the end of each segment. The goal is not perfection---we are simply interested in understanding how you approach realistic Excel tasks.

\paragraph{Instructions for Both Tasks.} When you are ready to begin, please inform the experimenter. You may use any standard Excel features during the task. The interface will also include a side panel that displays system-generated recommendations (depending on condition). You are encouraged to explore and apply these recommendations during the task.
\end{document}